\documentclass[twocolumn,showpacs,preprintnumbers,amsmath,amssymb]{revtex4}

\usepackage{graphicx}
\usepackage{dcolumn}
\usepackage{bm}
\usepackage{color}

\definecolor{red}{rgb}{1,0,0}
\newcommand{\beeq}{\begin{equation}}
\newcommand{\eneq}{\end{equation}}
\newcommand{\beeqa}{\begin{eqnarray}}
\newcommand{\eneqa}{\end{eqnarray}}

\begin{document}

\preprint{APS/123-QED}

\title{Minimal charge gap in the Ionic Hubbard Model} 

\author{Krunoslav Po\v{z}gaj\v{c}i\'{c}}
 \email{kpozga@lusi.uni-sb.de}
\author{Claudius Gros}%
\affiliation{%
Universit\"{a}t des Saarlandes, Institut f\"{u}r Theoretische Physik\\
}%

\date{\today}

\begin{abstract}							%
We study the ionic Hubbard model at temperature $T=0$ within the
mean-field approximation and show that the charge gap does
not close completely at the ionic-band insulator to 
antiferromagnetic 
insulator transition, contrary to previous expectations.
Furthermore, we find a new intermediate phase
for on-site repulsions $U>U_c$ for different lattices and calculate
the phase diagram for the ionic Hubbard model with
alternating $U$, corresponding to a Cu-O lattice.
\end{abstract}								%

\pacs{Valid PACS appear here}
\maketitle

%
%
%
%
\section{\label{sec:intro}Introduction}
The generalization of the Hubbard model with different on-site 
energies on the neighboring sites has been 
named the ionic Hubbard model (IHM),
%
%
\begin{eqnarray}        %
\label{H_IHM}
H&=&t\sum\limits_{<i,j>,\sigma}
c^{\dagger}_{j,\sigma}c_{i,\sigma}^{\phantom{\dagger}}
  +\frac{U}{2}\sum\limits_{i,\sigma} n_{i,\sigma} n_{i,-\sigma}
\\ \nonumber &&\qquad\quad  
+\,E_0\sum\limits_{i,\sigma}(-1)^{|i|} n_{i,\sigma},
\end{eqnarray}          %
%
%
which is characterized by a nearest-neighbor hopping
amplitude $t$, an on-site Coulomb-repulsion $U$
and a site-dependent on-site energy $\pm E_0$.
Here $|i|=$even/odd respectively for a A- and a B- site on a
bipartite lattice.

The IHM is used in two contexts:
(a) For the description of the neutral-ionic transition
(NIT) \cite{Torrance1} in organic mixed-stack charge-transfer
(CT) crystals.  The stacks in CT crystals form 
quasi-1d insulating chains with alternating
donor and acceptor molecules. 
The charge on the acceptor
$\rho$ and on the donor $-\rho$ characterize 
the crystal state. For $\rho<0.5$ the crystal
is said to be neutral-like, otherwise 
it is ion-like. 
The transition from one region into 
the other due to the change of temperature
or pressure is called NIT.
(b) The IHM has also been used in the context of 
ferroelectrics and superconductivity
in the transition-metal oxides \cite{Ishi1}. 
It has been argued \cite{Ishi1}
that the influence of the underlying lattice on the 
electronic system could be large in the critical region, leading
to a non-linear electronic polarizability.

The phase diagram of (\ref{H_IHM}) has been discussed previously
by several authors. Oritz {\it et al.} did find a single
phase transition in a mean-field study using a
one-dimensional density of states \cite{Ortiz}.
The transition takes place between paramagnetic and
the antiferromagnetic state. The point where it takes place 
has a special role with respect to the electrical conductivity. 
Whereas the system is a
semimetal in this point, in other points of the phase 
diagram it is an
insulator. The possibility of a semimetallic transition point
involving antiferromagnetism has also been discussed in a
LDA-study \cite{Leuken}.
 
Gidopoulos {\it et al.} performed a mean-field decoupling for
two-dimensional bipartite square and honeycomb lattices \cite{Gidopoulos}.
They did find two special values of $U$ for a given
energy-alternation $E_0$. The first denotes the phase-transition 
where the magnetic order sets in and at the second special value
for the one-site repulsion $U$ the charge-gap
for one spin species would presumably close.
In what follows, we'll show that there are actually two
phase transitions in the mean-field
decoupling scheme.
Furthermore, it will be shown that the solution where the gap for 
one spin species vanishes is thermodynamically not stable,
the charge-gap consequently does not close at any point in
the phase diagram.

\begin{figure}[htb]								%
\includegraphics[width=8cm,height=6cm]{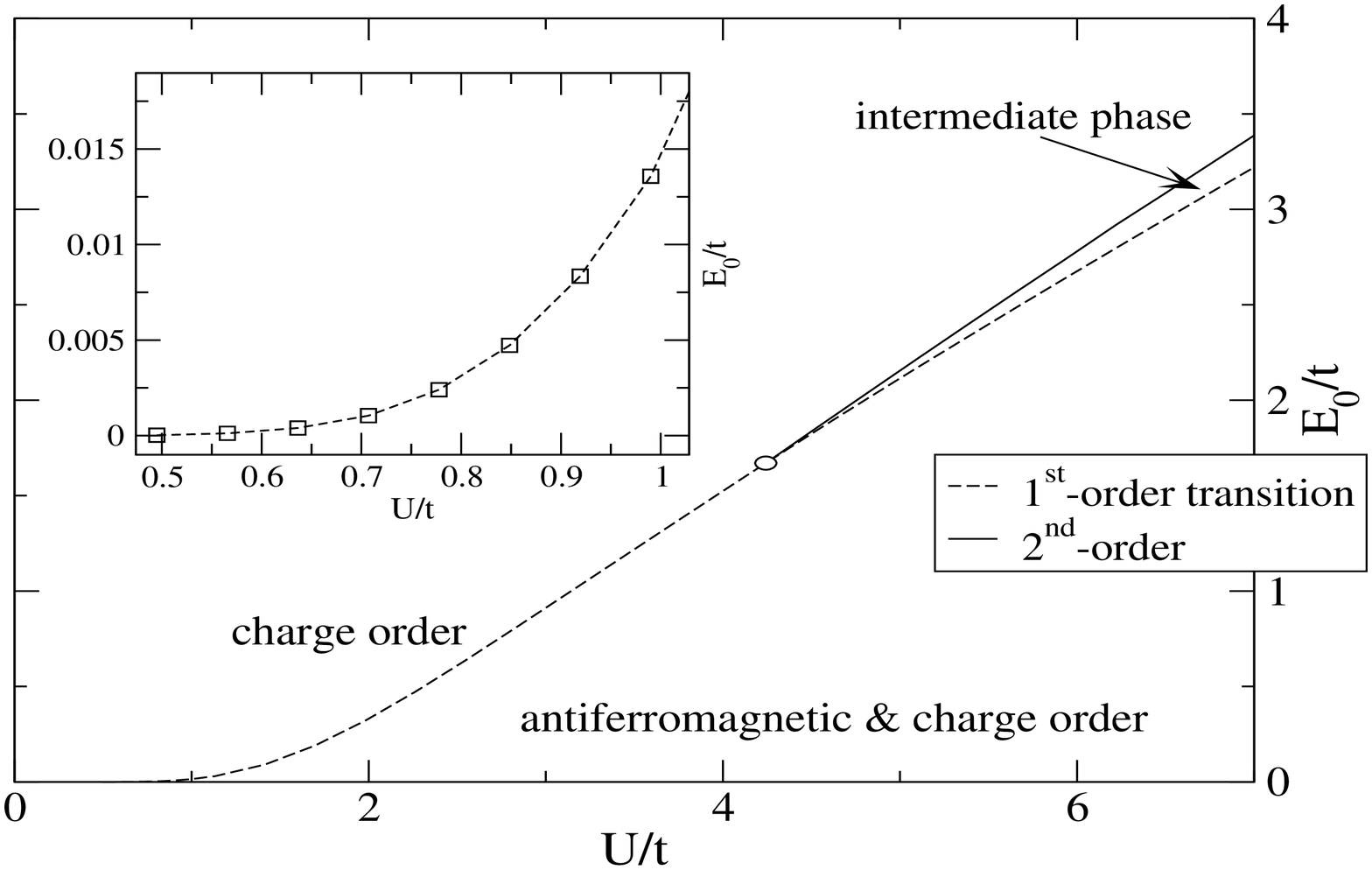}
\caption{\label{fig:PD_1d} 
Phase diagram for the 1d DOS. For $U<U_c$ 
the intermediate phase (I) is thermodynamically not stable and 
the transition from
antiferromagnetic (AF+CO) to paramagnetic phase (CO) 
is of the 1st order. The $U>U_c$ region has a stable (I)-phase. 
The transition between the two magnetic
phases (AF+CO and I) is of the
1st order, the (I)-phase goes into the (CO)-phase through
the 2nd order phase transition. The inset shows asymptotic behavior
of the phase border for small $U$.}
\end{figure}								%

%
%
%
%
\section{\label{sec:order_par}Order Parameters}

Application of the mean-field decoupling
%
%
\beeq
 n_{i\uparrow}n_{i\downarrow}\quad \rightarrow\quad
 \sum\limits_{\sigma}\langle n_{i,-\sigma}\rangle n_{i,\sigma}
 -\langle n_{i,\uparrow}\rangle \langle n_{i,\downarrow}\rangle
\eneq			%
%
%
leaves us with the system of equations
%
%
\beeq			%
\Delta_{\sigma}=-\int\limits_{0}^{\infty}d\epsilon D(\epsilon)
                                        \frac{E_{0}+U\Delta{-\sigma}}
					{\sqrt{\epsilon^2+(E_0+U\Delta_{-\sigma})^2}}
\label{selfT0_1}
\eneq			%
%
%
for the two order parameters $\Delta_\sigma$ defined
through the respective spin-densities 
$n_{A/B\sigma}= 1/2 \mp \Delta_\sigma$ on A- and B-sites.
$D(\epsilon)$ is a free-particle density of states (DOS). In what follows,
1d-DOS will be used, but qualitatively the same results have been obtained
for the flat and the semicircular DOS.
Quasi-particle states are accommodated in four bands with dispersion
%
%
\beeq			%
\label{eq:disp}
\lambda_{\alpha\sigma}(\epsilon)=
\frac{U}{2}+\alpha\sqrt{\epsilon^2+(E_0+U\Delta_\sigma)^2},
\qquad \alpha=\pm 1.
\eneq			%
%
In the ground-state the $\alpha=-1$ bands are filled.
The thermodynamic stability of a possible self-consistent
solution of (\ref{selfT0_1})
is determined, at $T=0$, by the total energy 
%
%
\begin{eqnarray}        %
\label{energy_1d_T0}
\frac{E}{N}&=&\sum\limits_{\alpha,\sigma}\int\limits_{0}^{\infty}
 d\epsilon\, D(\epsilon)\,\lambda_{\alpha\sigma}(\epsilon)
 \,\Theta(\epsilon_F-\lambda_{\alpha\sigma}(\epsilon))
\\ \nonumber && \qquad\quad 
 -\,\left( U/4 + U\Delta_\uparrow\Delta_\downarrow.
\right)
\end{eqnarray}          %
%
%
Above equations are valid for the half-filled system.
Use of the particle-hole symmetry
%
%
\beeq			%
c_{j\sigma}|FB\rangle =(-1)^{|j+1|}d^{\dagger}_{j+1,-\sigma}|0\rangle_h
\eneq			%
%
where $|FB\rangle$ denotes a full band and $|0\rangle_h$ 
a hole vacuum \cite{Mahan}, shows that $\epsilon_F=U/2$. 
Together with the dispersion of the bands, this implies 
that the number of each spin species is the same.

%
\begin{figure}[thb]							%
\includegraphics[width=7cm,height=5cm]{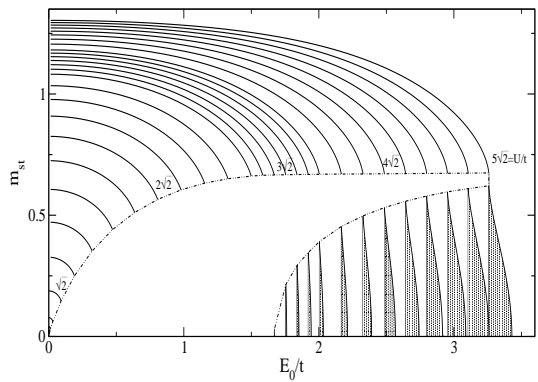}
\caption{\label{fig:AF_order_par} 
Staggered magnetization curves for different $U$'s. 
Blank, inclosed region contains jumps of the $m_{st}$. Shaded regions
are placed under the magnetization curves of the (I)-phase.  }
\end{figure}								%
%

%
%
%
%
\section{\label{sec:phase_dia}Phase Diagram}

The phase diagram of the IHM is shown on Fig.~\ref{fig:PD_1d}. Regarding
the $U$-axis, it has two separate regions. For 
$U<U_c\approx 4.25t$ one finds two phases: 
the charge-ordered (CO)-phase and a mixture of the charge ordered and the
antiferromagnetic phase (AF+CO). In the $U>U_c$ region there are three phases.
Beside the (CO) and the (AF+CO)-phases, 
we find another phase which we will call intermediate (I)-phase. 
The (I)-phase has the same order parameters as (AF+CO)-phase. 

The phases shown in Fig.~\ref{fig:PD_1d} can be observed clearly
in the curves for the staggered magnetization 
$m_{st}=|\Delta_\downarrow-\Delta_\uparrow |$ shown in
Fig.~\ref{fig:AF_order_par} as a function of $E_0$ and
various values of $U$.  We notice a jump in the curves for
all values of $U$, indicative for the 1st order phase transition. 
The jumps define a region of $m_{st}$ for which
the system is unstable, illustrated by the 
light shaded area in Fig.~\ref{fig:AF_order_par}.
When $U>U_c$ there is no direct transition from 
the low-$E_0$ (AF+CO)-phase to the high-$E_0$ (CO)-phase which is
characterized by $m_{st}=0$. The
(AF+CO)-phase now shares a border with (I)-phase.
The jump in the staggered magnetization vanishes
asymptotically as $U$ becomes large. 
The transition from (I) to (CO) is, on the other hand, of the 2nd order. 
From the magnetization curves it is also transparent that the
width of (I) rises as one increases $U$ 
(illustrated by the dark-shaded areas
in Fig.~\ref{fig:AF_order_par}).

\begin{figure}[htb]							%
\includegraphics[width=8cm,height=5cm]{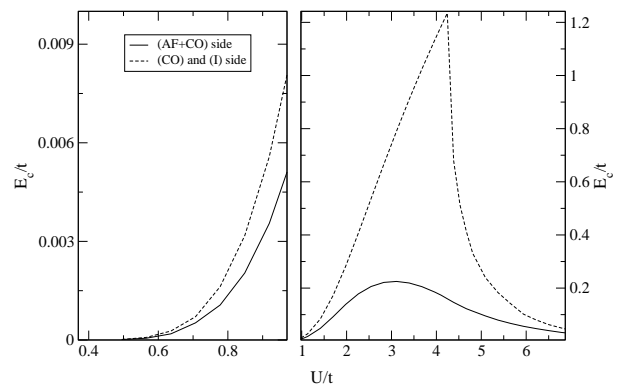}
\caption{\label{fig:Charge_gap} Charge gap as a function of $U$ on the
1st order transition point (AF+CO)$\rightarrow$(I) for $U>U_c$ and
(AF+CO)$\rightarrow$(CO) for $U<U_c$. On the left plot we zoomed in the curves
showing behavior of the gaps in the 
region of the parameter space where the $(AF+CO)$ phase shrinks rapidly.}
\end{figure}

The first-order nature of the transition between the
(AF+CO)-phase realized for low values of $E_0$ and
the (CO) and (I)-phase respectively shows up in a jump of the charge-gap,
as shown in Fig.~\ref{fig:Charge_gap}.
The discontinuity in the charge-gap is manifested through two different
curves of the gap as a function of $U$ for the two-sides of the
transition respectively.
%
%
%

\begin{figure}[htb]							%
\includegraphics[width=7cm,height=5cm]{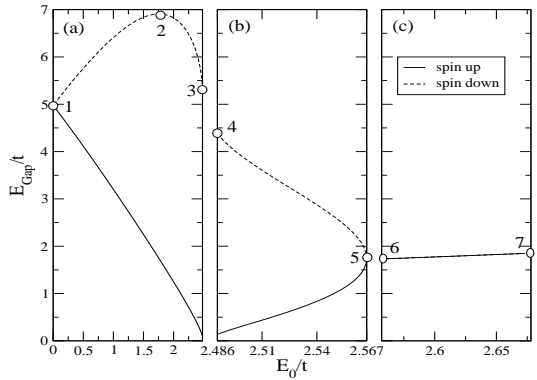}
\caption{Evolution of the
gaps for each spin direction as a function of $E_0$, for fixed
$U=4\sqrt{2}~t$.
The smaller gap is equivalent to the charge gap.
(a) (AF+CO)-phase, the charge gap decreases.
(b) (I) state, the charge gap increases. 
(c) (CO)-phase, the gap for both spin directions is the same.
} \label{fig:gap_evolution}
\end{figure}

A difference between (I) and (AF+CO) is in the behavior 
of the charge gap as a function of $E_0$, see
Fig.~\ref{fig:gap_evolution}. 
The transition point between the (AF+CO) and (I)-phase is characterized by a
minimum in the charge gap, see Fig.~\ref{fig:AF_order_par}. 
Even though the gap is close to zero, it remains finite. 
In the (AF+CO)-phase 
(Fig.~\ref{fig:gap_evolution}a) the charge gap, 
as a function of $E_0$, decreases while 
it increases in the (I)-phase (Fig.~\ref{fig:gap_evolution}b). 
As expected, this increase proceeds in the (CO)-phase
(Fig.~\ref{fig:gap_evolution}c).

%
%
%
\begin{figure*}[htb]							%
\includegraphics[width=8cm,height=6cm]{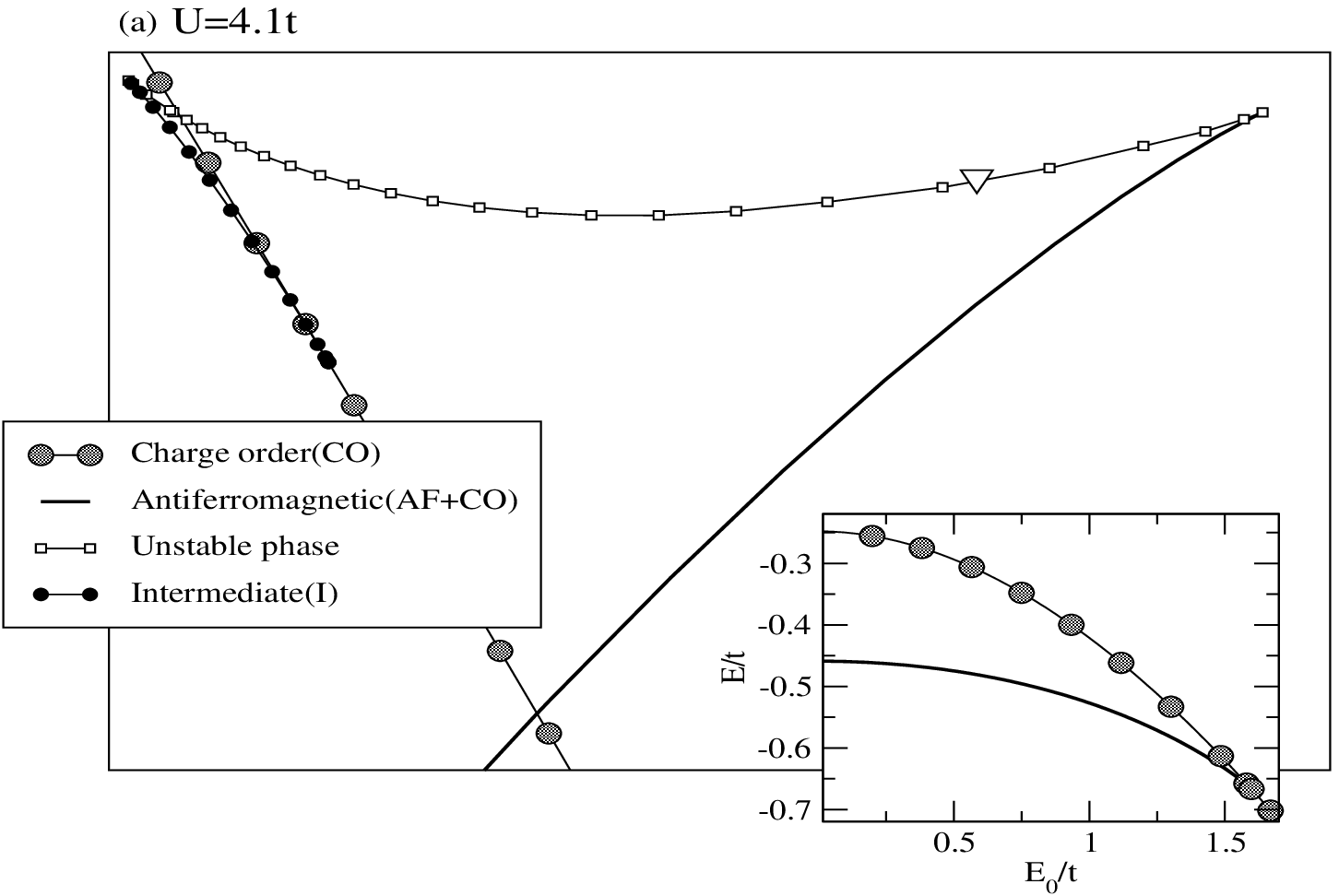}
\includegraphics[width=8cm,height=6cm]{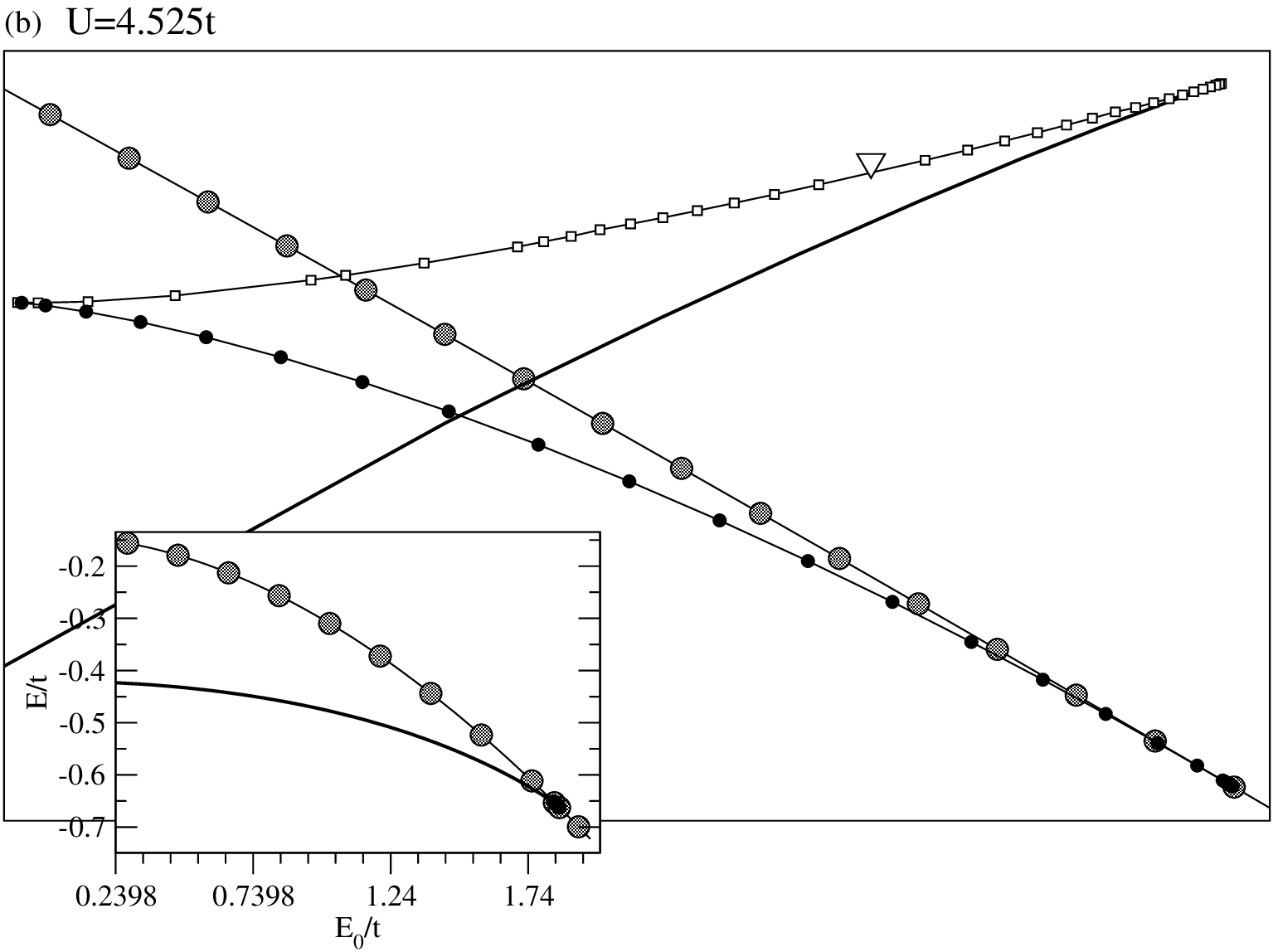}
\caption{\label{fig:thermo_stability}
 (a) Energy of the various states as
a function of $E_0$. 
$U=2.9\sqrt{2}~t<U_c$ is chosen such that the (I)-phase is unstable.
The main panel is blowup of the energies shown in the inset,
transformed by the subtraction of a suitably chosen straight 
so that the particular transitions can be easier visualized. 
The large  empty triangle denotes the position of
the vanishing charge gap in the thermodynamically unstable solution.
 (b) $U=3.2\sqrt{2}~t>U_c$ so that the part of (I)-phase is stable.
}
\end{figure*}								%
%

%
%
%
%
\section{\label{sec:thermo}Thermodynamic stability}

A self-consistent solution of Eq.~(\ref{selfT0_1})
with finite values for the
order parameters $\Delta_\sigma$
is not a guarantee that the corresponding phase is
indeed realized in the system. The condition for the stability of a phase on
$T=0$ is that it has the minimal energy with respect to the energies
of other phases. This
fact is important in the case of (AF+CO)-(I) transition. There, the (AF+CO)
and the (I)-solutions show a hysteresis effect and the ground state is
determined on the basis of the minimum energy principle.

The examination of the energy space for the case $U<U_c$ is shown on
Fig.~\ref{fig:thermo_stability}(a). On it one sees the energy 
Eq.~(\ref{energy_1d_T0}) for the four different solutions 
which solve the self-consistency Eq.~(\ref{selfT0_1})
for the order parameters $\Delta_\sigma$. They are
the three solutions (AF+CO,line), (CO, large filled circles) and
intermediate phase (I, small filled circles) which we have already
discussed, together with the fourth solution, which is always
unstable (empty squares). 
The energy of the (I)-phase is larger than the one of the (AF+CO)-phase 
in the whole interval of
its existence. (I)-phase is thus unstable. Crossing of the energies of
(AF+CO) and (CO)-phase discloses an underlying 1st order phase transition. 
The lower inset shows energies of all phases in the whole $E_0$ interval.

For $U=3.2\sqrt{2}~t>U_c$ the results are given in 
Fig.~\ref{fig:thermo_stability}(b). 
Part of the (I)-phase is now stable. 

Stable solutions
have been obtained by iterating Eq.~(\ref{selfT0_1}) directly
and the unstable solutions 
by fixing one of $\Delta_\sigma$'s and allowing $E_0$ to change. 
The fixed order-parameter $\Delta_\sigma$
has been chosen such that it is in the range which is not covered by the
stable solutions. We have also tried to use fixed $U$, but the 
self-consistency map, for the parameters we examined, didn't turn out to
be attractive and thus useful. Limiting cases:(1) $E_0=0$ and (2) $U=0$
assure us that in this limits we found all solutions. Thus, possible
solutions with the weakly attractive self-consistency map are a possibility
in the intermediate range of parameters. We cannot exclude possibility
of their existence, but due to a few self-consistency arrangements we
find them improbable.

The charge gap vanishes when
$\Delta_{\sigma}=-E_0/U$ for one spin species and
$\Delta_{-\sigma}=0$ for the other species.
The energy of this solution
is given in Fig.~\ref{fig:thermo_stability}(a) and
Fig.~\ref{fig:thermo_stability}(b) and is denoted by
an empty triangle. 
It turned out to be unstable in the calculations done. Furthermore,
from the performed calculations (see Fig.~\ref{fig:thermo_stability})
we see that the vanishing-gap solution has always (AF+CO) and (CO) 
companions with the smaller energy. This leads us to extend the conclusions 
obtained from our results to the whole parameter region, including
interval $U/t<0.5$ which we haven't investigated numerically. Thus,
the charge gap closes only in the point $E_0=U=0$.

Finally, we would like to mention that the three stable phases
(CO), (I) and (CO+AF) are characterized by different distribution
of the charge densities
$n_{A/B\sigma}= 1/2 \mp \Delta_\sigma$ on A- and B-sites.
We find: $\Delta_\uparrow=\Delta_\downarrow$ for the (CO) phase,
$\Delta_\uparrow\ne\Delta_\downarrow$ and
$\Delta_\uparrow \Delta_\downarrow>0$ for the (I) phase and 
$\Delta_\uparrow \Delta_\downarrow<0$ for the (AF+CO) phase.
%
%
%
%
%
\section{\label{sec:AlterU}Alternating $U$ Ionic Hubbard Model (AIHM)}
A  natural generalization of the ionic Hubbard model 
is the model where on-site Coulomb
interaction on the atoms $A$ and $B$ is not the same. 
The Hamiltonian of the new system may be written as
%
%
\beeq			%
\label{eq:AIHM}
H_{AIHM}=H_{IHM}+\sum\limits_i dU(-1)^{|i|} n_{i\uparrow}
n_{i\downarrow}
\eneq			%
%
%
For simplicity we consider here the extreme case where 
the Coulomb interaction disappears on site $A$. 
This implies $dU=-U$ and
a value of $2U$ for the Coulomb repulsion on site $B$.
This model is than equivalent to the Cu-O lattice model 
\cite{Ortiz} with correlated
$B$-sites (Copper, lower on-site energy) and uncorrelated
$A$-sites (Oxygen, higher on-site energy).

Let's write $E_0=U/2+e_0$ and $\mu=U/2+\mu'$, 
where $\mu$ is the chemical potential for the half-filled
system. The transformation 

$$
E_0=U/2+e_0\quad\rightarrow\quad E'_0=U/2-e_0
$$ 
and 
$$
\mu=U/2+\mu'\quad\rightarrow\quad \mu=U/2-\mu'
$$ 
leaves us with an equivalent Hamiltonian.
This can be shown by performing the canonical particle-hole transformation 
$c_{j,\sigma}\rightarrow (-1)^{|j|}d^{\dagger}_{j,-\sigma}$. 
It yields
%
%
\begin{eqnarray}	%
H=\sum\limits_{<i,j>,\sigma}td^{\dagger}_{j,\sigma}d_{i,\sigma}
  +\sum\limits_{i} U(1-(-1)^{|i|}) n^h_{i\uparrow} n^h_{i\downarrow} 
  \nonumber \\
  +\sum\limits_{i,\sigma}(-1)^{|i|}(\frac{U}{2}+e_0) n^h_{i,\sigma} \nonumber\\
  -\sum\limits_{i,\sigma}(-1)^{|i|}(\frac{U}{2}+\mu') n^h_{i,\sigma}
+2N\mu' 
\label{eq:holeH}
\end{eqnarray}		%
%
%

The difference with respect to the original
Hamiltonian (\ref{eq:AIHM})
lies in the constant term $2N\mu'$. This, together with the
fact that the order parameter operators preserve the same
form in the hole picture, implies that the phase 
diagram of (\ref{eq:AIHM}) is symmetric with respect 
to the line $E_0=U/2$ in the parameter $E_0$ and for
$dU= -U$.

The self-consistency equation, derived under the assumption 
that the ferromagnetic order parameter vanishes, 
is given by
%
\beeq			%
\Delta_{\sigma}=-\int\limits_{0}^{\infty}d\epsilon D(\epsilon)
			\frac{B_\sigma}
			{\sqrt{\epsilon^2+B_\sigma^2}}~,
\label{selfT0_2}
\eneq			%
%
%
where $B_\sigma=E_0+U\Delta_{-\sigma}+dU/2$.

The energy spectrum has a form
$\lambda_{\pm,\sigma}(\epsilon)=A_\sigma\pm\sqrt{\epsilon^2+B_\sigma^2}$
 where $A_\sigma=U/2+dU\Delta_{-\sigma}$.
\par
\begin{figure}[htb]							%
\includegraphics[width=7cm,height=5cm]{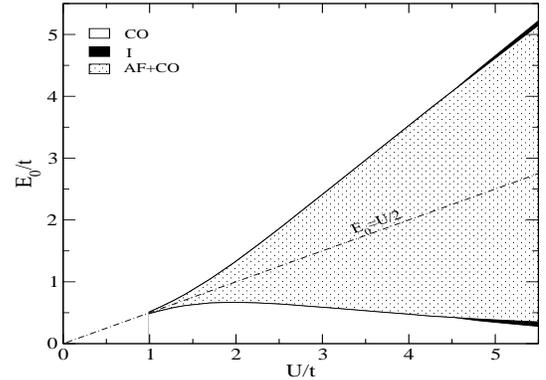}
\caption{\label{fig:PD_alterU} 
Phase diagram of the half-filled AIHM
with the 1d density-of-states.
It contains three phases. The charge ordered 
(CO)-phase is given by a white area. 
The dotted area represents a mixture of
the antiferromagnetic and the charged phase (AF+CO) 
and the black region covers a parameter range where intermediate 
(I)-phase appears. The symmetry of
the phase diagram is denoted by the dash-dot line.}
\end{figure}

In Fig.~\ref{fig:PD_alterU} we present the
phase diagram of the AIHM. It contains the same phases as the 
IHM phase diagram. The dash-dot line
in Fig.~\ref{fig:PD_alterU} indicates the symmetry axis. 
On it the system is antiferromagnetic with the
vanishing charge-order parameter (AF).
In the region around the symmetry line the system is in the (AF+CO)
phase which, for $U<U_c$, makes transition into the pure charge-ordered
state. For $U>U_c$ the transition takes place into the (I)-phase
and than into the (CO)-phase.
%
%
%
%
\par
Let's define a singlet gap as a minimal energy needed for a transfer 
of the electron in an empty state without a spin-flip and
the triplet gap as a minimal energy of the transition where the spin 
is flipped and the total $S_z$ changes. In the mean-field formulation 
this gaps are given by $\Delta_s=2|B_\sigma|$ and
$\Delta_t=\left|A_\uparrow+|B_\uparrow|+|B_\uparrow|-A_\downarrow\right|$.
\par
\begin{figure}[htb]							%
\includegraphics[width=8cm,height=8cm]{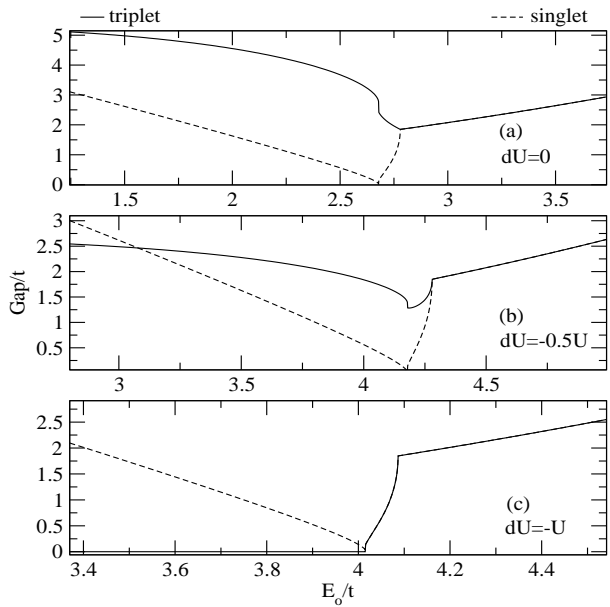}
\caption{\label{fig:single_triple_gap} Singlet (dashed curve) and triplet
(full curve) gaps as a function of the on-site 
chemical potential amplitude $E_0$ for $U=6t$.
(a) $U$ is the same for both atom species. (b) $U_A<U_B$
(c) $U$ on site A vanishes.
}
\end{figure}

A distinct property of the AIHM is the closing
of the triplet gap in the (AF+CO) phase. This is illustrated
in Fig.~\ref{fig:single_triple_gap}. IHM case with a uniform $U$
is shown on Fig.~\ref{fig:single_triple_gap}(a). In
Fig.~\ref{fig:single_triple_gap}(c) we see that the triplet gap
vanishes for $U_A=0$ and $U_B=2U$. 
%
%
%
%
A necessary (but not sufficient) condition for the vanishing of the triplet gap
can be derived from Eq.(\ref{selfT0_2}) and expressed in the
form $\Delta_\uparrow\Delta_\downarrow<0$.
This is the case in the (CO+AF) phase.
$\Delta_\uparrow\Delta_\downarrow>0$ and the triplet gap is 
non-vanishing as can be seen in Fig.~\ref{fig:single_triple_gap}.
The exception to the condition $\Delta_\uparrow\Delta_\downarrow<0$ for
disappearance of the triplet gap is a solution 
$\Delta_\uparrow=\Delta_\downarrow=0$ which exists for $E_0=\frac{U}{2}$.
%
%
%
%
%
%
%
%
\section{\label{sec:discussion}Discussion}
Previous mean-field studies of the ionic Hubbard model did find
a vanishing charge gap at the transition
point between antiferromagnetic and paramagnetic state
\cite{Ortiz,Gidopoulos}, in contrast to our result of a
minimal charge gap. Presented data has been obtained 
for a 1d one-electron density-of-state. For comparison
we have carried out 
calculation also for the flat and the semicircular density-of-states.
The results changed only quantitatively, all major features discussed 
previously remain valid.

It is interesting to compare with the 1d-ionic Hubbard model,
which was studied by Resta and Sorella \cite{Resta95} using
the boundary-condition-integration technique \cite{Gros92}, 
by Brune {\it et al.} using DMRG \cite{Brune1} and by Wilkins
and Martin using QMC \cite{Wilkens}. Fehske {\it et al.}
considered the dynamical IHM coupled to phonons
\cite{Fehske}.

Fabrizio {\it et al.}, in a Bosonization-study \cite{Fabrizio},
proposed the existence of a dimerized intermediate phase
in the 1d ionic Hubbard model.  DMRG- and QMC-results indicate
\cite{Brune1,Wilkens} that this bond-ordered (dimerized)
insulating-phase extends to arbitrarily large values of $U$, due to
the instability of the 1d Mott-Hubbard insulator towards dimerization.
DMRG studies of the 1d-model \cite{Takada,Brune1}, 
found indications of a `strange' metallic point with a
finite charge gap at the transition
from the band-insulator to the correlated insulating-state.
We may speculate, that our mean-field result
of two distinct phase transitions reflects, on the other hand,
the occurance of
magnetic long-range order possible in dimensions larger than
one. We note, in this context, that the intermediate phase (I)
evidenced in Fig.~\ref{fig:PD_1d} and Fig.~\ref{fig:PD_alterU} 
does not show spontaneous dimerization.\\
%
%
%
%
\section{\label{sec:conclusion}Concluding remarks}
The mean-field decoupling scheme is an approximation to the full evaluation of
the Hamiltonian. Its validity is restricted to the small U limit
and large dimension where it is qualitatively correct \cite{Dongen}.
We have found, however, that the stability-analysis 
for the various solutions possible for the ionic
Hubbard model is highly non-trivial. 
Here we have presented two new results:
(i) A non-vanishing charge gap for all parameters
and (ii) the existence of two distinct phase transitions for larger
values of $U$. 
%
%

%
%
\end{document}